\documentclass{article}
\usepackage[a4paper, total={6in, 8in}]{geometry}

\usepackage[most]{tcolorbox}

\newtcolorbox{greybox}{
  colback=gray!10,
  colframe=gray!50,
  boxrule=0.4pt,
  arc=1mm,
  left=7pt,
  right=7pt,
  top=7pt,
  bottom=7pt
}

\usepackage{lipsum}
\usepackage{endnotes}
\usepackage{natbib}
\usepackage{graphicx}
\usepackage{moreverb,url}
\usepackage{xcolor}
\usepackage{lipsum}  
\usepackage{footmisc}
\usepackage{enumitem}
\usepackage{booktabs}

\usepackage{amsmath}
\usepackage{amssymb}
\usepackage{authblk}

\usepackage{tabularx}
\usepackage{multirow}
\usepackage{hyperref} % temporary, for easier navigation

\usepackage{gensymb}

\usepackage{float}
\usepackage[normalem]{ulem}

\usepackage[latin1]{inputenc}

\usepackage{caption}
\newcommand{\citeauthorpos}[1]{{\citeauthor{#1}'s}}

\newcommand{\citepos}[1]{{\citeauthorpos{#1}~\citeyearpar{#1}}}

\title{%
  \fontsize{22}{26}\selectfont
  Methods for pitch analysis \\in contemporary popular music:\\
  multiple pitches from harmonic tones \\in Vitalic's music%
}

\author[1,2]{\vspace{1cm} Emmanuel Deruty}
\author[3]{Maarten Grachten}
\author[2]{David Meredith}
\author[4]{Pascal Arbez-Nicolas}

\author[2]{\\Andreas Hasselholt J{\o}rgensen}
\author[2]{Oliver S{\o}nderm{\o}lle Hansen}
\author[2]{\\Magnus Stensli}
\author[2]{Christian N{\o}rk{\ae}r Petersen}

\affil[1]{Sony Computer Science Laboratories, Paris, France}
\affil[2]{Department of Architecture, Design and Media Technology, Aalborg University, Aalborg, Denmark}
\affil[3]{Contractor for Sony Computer Science Laboratories, Paris, France}
\affil[4]{Citizen Records, Dijon, France}

\date{}

\begin{document}
\maketitle

\begin{greybox}
Published version: Deruty, E., Grachten, M., Meredith, D., Arbez-Nicolas, P.,
J{\o}rgensen, A. H., Hansen, O. S., Stensli, M.,
\& N{\o}rk{\ae}r Petersen, C. (2026).
Methods for pitch analysis in contemporary popular music:
Multiple pitches from harmonic tones in Vitalic's music.
\textit{Journal of the Audio Engineering Society}, 74, 270--287.
\url{https://doi.org/10.17743/jaes.2022.0264}
\end{greybox}

\vspace{0.5cm}

\begin{abstract}

\textbf{Aims.} This study suggests that the use of multiple perceived pitches arising from a single harmonic complex tone is an active and intentional feature of contemporary popular music. The phenomenon is illustrated through examples drawn from the work of electronic artist Vitalic and others.
\textbf{Methods.} Two listening tests were conducted: (1) evaluation of the number of simultaneous pitches perceived from single harmonic tones, and (2) manual pitch transcription of sequences of harmonic tones. Relationships between signal characteristics and pitch perception were then analyzed.
\textbf{Results.} The synthetic harmonic tones found in the musical sequences under study were observed to transmit more perceived pitches than their acoustic counterparts, with significant variation across listeners. Multiple ambiguous pitches were associated with tone properties such as prominent upper partials and particular autocorrelation profiles.
\textbf{Conclusions.} Harmonic tones in a context of contemporary popular music can, in general, convey several ambiguous pitches. The set of perceived pitches depends on both the listener and the listening conditions.

\end{abstract}

\clearpage
%Head 1

\section{Introduction}\label{sec:introduction}

The human ability to perceive distinct pitches simultaneously from a single harmonic complex tone -- referred to here as \emph{tonal fission} -- has long been debated \citep{turner1977ohm} and demonstrated using artificial stimuli \citep{shepard1964circularity,moore2006frequency,yost2009pitch}. However, tonal fission has generally not been considered significant in everyday musical listening, due to the specific conditions thought necessary for it to occur \citep{plomp1976aspects,dixonward1970}. Harmonic complex tones produced by musical instruments are typically considered to transmit a single pitch, corresponding to the fundamental frequency ($f_0$) -- the lowest harmonic of the tone.

Nevertheless, the phenomenon plays a documented role in music. The deliberate elicitation of multiple pitches from a single quasi-harmonic tone appears in non-Western traditions such as overtone singing \citep{bloothooft1992acoustics}, as well as in contemporary classical and jazz music, where it is referred to as \emph{multiphonics} \citep{fox2020art,fallowfield2020cello} and employed as a compositional or expressive technique.

In contemporary popular music (CPM), as defined by \citet{deruty2022development}, the phenomenon of tonal fission has remained largely undocumented. Only recently has it been identified as a stylistic element in the works of contemporary electronic music producers such as Hyper Music \citep{deruty2022melatonin,deruty2025insights, deruty2025primaal} and Vitalic \citep{deruty2025vitalictemperament}. These studies make two key observations. First, the ubiquitous use of synthesizers and effects processors gives artists a high degree of control over sound, allowing them to create quasi-harmonic tones that are ambiguous in pitch content. Second, \citet{deruty2025primaal,deruty2025vitalictemperament} find that the artists they studied use tonal fission and pitch ambiguity \textit{deliberately} as compositional tools -- traits that may be considered idiomatic to the technology in the sense defined by Huron  \citep{huron2009characterizing}.

Whereas prior work by \citet{deruty2025primaal,deruty2025vitalictemperament} draws on producers' perspectives and signal analysis to question the link between a harmonic tone and a single pitch, the present study approaches this issue through listening tests focused on tonal fission. These tests use excerpts from the electronic music producer Vitalic (with whom the authors were in contact during the development of the study), alongside samples representative of other genres, including distorted electric guitar and TR-808-style bass. The results suggest that single quasi-harmonic complex tones in these contexts can give rise to multiple pitch sensations, with perceived pitches varying across listeners.

The paper is structured as follows. Section~\ref{sec:background} presents the context of the study, and Section~\ref{sec:methods} outlines the methods. Sections~\ref{sec:listening1} and~\ref{sec:listening2} detail the two listening tests and their respective results. Section~\ref{sec:signal} explores links between a tone's signal and the phenomenon of tonal fission. Section~\ref{sec:comments} compiles listener comments to provide further insight into the test results. Broader perspectives are discussed in Section~\ref{sec:discussion}, and conclusions are presented in Section~\ref{sec:conclusion}. Supplementary material with audio examples is available at:

\vspace{.3cm}

\url{http://vhtmp-suppl-mat.s3-website.eu-west-3.amazonaws.com/}

%%%%%%%%%%%%%%%%%%%%%%%%%%%%%%%%%%%%%%%%%%%%%%%%%
%% BACKGROUND
%%%%%%%%%%%%%%%%%%%%%%%%%%%%%%%%%%%%%%%%%%%%%%%%%

\newpage
\section{Background}\label{sec:background}

This section introduces the notion of \textit{tonal fission}, in contrast to \textit{tonal fusion}. It outlines how Vitalic's music production deliberately employs tonal fission for expressive purposes and defines several useful concepts related to pitch perception.

\subsection{Tonal fusion and fission}

The term \emph{tonal fusion} refers to the tendency for certain combinations of tones to be perceived as a single auditory image \citep{stumpf1890tonpsychologie, dewitt1987tonal}. Typically, tones separated by intervals with strong harmonic relationships are more likely to produce tonal fusion across listeners. In a distinct context, the term \emph{fission} has been used to describe the perception of a single tone sequence as originating from two or more sources \citep{vannoorden1975coherence, bregman1994auditory, moore2012properties}.

The samples examined in this study exhibit \emph{fission arising from a single tone}, a phenomenon referred to here as \emph{tonal fission}. It relates to the well-documented ability of humans to perceive multiple distinct pitches simultaneously from a single (quasi-)harmonic complex tone -- a topic that has long been the subject of research and debate \citep{turner1977ohm, mersenne1636harmonie, rameau1750demonstration, helmoltz1885sensations, plomp1964ear, plomp2001intelligent}. \citet{dewitt1987tonal} also suggest a conceptual link between tonal fusion and tonal fission, hypothesizing that the pitch ambiguity observed in the fusion of two harmonic tones could likewise arise from a single harmonic tone.

\subsection{Vitalic's perspective: tonal fission as a musical parameter}\label{sec:Vitalic}

A notable aspect of the work of electronic music producer, Vitalic, is what he describes as `a melody inside the bass' \citep{musicradartech2014vitalicITV}, referring to a secondary melodic line embedded within a sequence of distinct quasi-harmonic tones. More broadly, Vitalic has confirmed in private interviews that he treats the number of perceived pitches arising from a single tone as a musical parameter, and that these pitches are often intentionally ambiguous. This approach is not unique to his practice, as suggested by studies involving other music producers and examples from the broader music market \citep{deruty2022melatonin, deruty2025insights, deruty2025primaal, deruty2025vitalictemperament, deruty2025vitalicnonharmonic}.

\subsection{Terminology}\label{subsec:terminology}

Regarding tones, we use the following terminology.

\begin{enumerate}%[noitemsep]

\item \emph{Partial}: the spectral representation of a single periodic sine wave, corresponding to what \citet{helmoltz1885sensations} describes as a \emph{constituent} or \emph{partial tone}.

\item \emph{Harmonic complex tone}: an ensemble of partials that are integer multiples of a fundamental frequency ($f_0$), as defined by \citet{helmoltz1885sensations} as a \emph{compound}. When the partial corresponding to $f_0$ is absent (`missing fundamental'), it may still be determined as the greatest common divisor of the partials' frequencies.

\item \emph{Overtone} or \emph{upper partial}: refers to any partial in a tone other than $f_0$. This aligns with Helmholtz's definition of \emph{upper partial tones} \citep{helmoltz1885sensations}.

\item \emph{Inharmonic tone}: a tone that is not harmonic. Tones can vary in their degree of inharmonicity, allowing for the identification of \emph{quasi-harmonic tones}, in which each partial of the least deviating harmonic series can be clearly associated with a partial of the original tone. In this case, $f_0$ may be evaluated as the fundamental frequency of the least deviating harmonic series \citep{rasch1982perception}.

\item \textit{Harmonics}: defined by \citet{helmoltz1885sensations} as the sinusoidal components (partials) of a harmonic complex tone, where the $n^{th}$ harmonic has a frequency $n$ times the fundamental. In the case of quasi-harmonic tones, the $n^{th}$ harmonic's frequency neighbours the $n^{th}$ multiple of the $f_0$ of the least deviating harmonic series. 

\end{enumerate}

\subsection{Pitch perception: spectral and temporal modeling}\label{subsec:modeling}

\citet{yost2009pitch} distinguishes between \textit{spectral} and \textit{temporal} models of pitch perception, whose main proponents are \citet{goldstein1973optimum} and \citet{licklider1951duplex}, respectively. In spectral modeling, pitch is extracted from spectral components that are \textit{resolved} by the auditory periphery -- that is, components whose frequency spacing exceeds the width of critical bands. In contrast, temporal modeling typically derives pitch from the autocorrelation of the signal. Depending on the tone, the spectral and temporal approaches may either converge or diverge.

In this paper, we apply simplified forms of both approaches to link perceived pitches with measurable signal characteristics. A simplified version of \textit{spectral modeling} is used to associate perceived pitches with individually audible upper partials. While \textit{temporal modeling} often relies on the highest autocorrelation peak beyond zero lag \citep{rabiner1976comparative}, we also consider frequency differences between adjacent partials as indicators of temporal structure.

\section{Methods}\label{sec:methods}

This section describes two methods used in the context of the study: audio source separation and equal-loudness-contour weighting. It then outlines the prior methodology that led to the present investigation.

\subsection{Source separation}\label{subsec:sourceseparation}

Signal analyses in this paper are based either on the original song versions or on audio files obtained through source separation using the X-UMX algorithm \citep{stoter2019open}. This algorithm separates audio into six categories: drums, guitar, bass, piano, vocals, and `other'. The sum of the separated files exactly reproduces the original.

Vitalic's bass tracks often differ from traditional electric bass, leading the algorithm to classify the lower bass spectrum as `bass' and the upper spectrum as `other'. Similarly, the lower range of harmonic synthesizers may also be categorized as `bass'. This ambiguity is reflected in some samples used in the study.

%This highlights an ambiguity between bass and synthesizers in Vitalic's music.

\subsection{Psychoacoustic weighting}\label{subsec:iso}

Signal analyses in this study incorporate equal-loudness-contour weighting, recognizing that humans are not equally sensitive to all frequencies \citep{fletcher1933loudness}. Various models exist that weight the power spectrum to reflect perception \citep{fletcher1933loudness,robinson1956re,skovenborg2004evaluation}. One such model is ISO226:2023 \citep{iso2262023}. We apply the ISO226:2023 50-phon equal-loudness contour before spectral analysis to produce results that better reflect what listeners actually hear.

\subsection{Prior analytical process}\label{subsec:analyticalprocess}

Seventy-two songs originating from seven Vitalic records released between 2005 and 2023 were imported into a Pro Tools session and subjected to critical listening. The focus of the listening was \textit{pitch}---not the high-level organisation of pitches such as chord sequence identification, but lower-level criteria, such as the complexity and nature of the tones carrying pitch or the compliance of pitches to equal temperament. Observations were sorted into six non-exclusive categories:

\begin{enumerate}[noitemsep]
    \item Presence of complex instrumental tones.
    \item Presence of complex vocal tones (e.g., vocoder).
    \item Tuning not consistent with equal temperament.
    \item Presence of continuous pitch trajectories.
    \item Presence of low pitch-strength instruments, e.g., drums.
    \item Presence of flanger/chorus resulting in additional pitch.
\end{enumerate}

The present study focuses on the first category. Of the 72 songs examined, 38 feature instrumental tracks based on tones that evoke more than one pitch. In most cases, these tracks are prominent and use such tones throughout the entire song. Therefore, while the study's conclusions do not apply to all of Vitalic's work, they are representative of over half his discography.

Table~\ref{table:extracts} lists the extracts used in this study. Nine extracts from Vitalic's music were selected for Listening Test 1 (Section~\ref{sec:listening1}), with three also used in Listening Test 2 (Section~\ref{sec:listening2}). The primary selection criterion was ensuring diversity within a small sample pool. Listening Test 1 also includes three monodic sequences from acoustic instruments, serving as a control group. One of these sequences was included in Listening Test 2, which also features a guitar power chord and a TR808 bass-style sample. These two samples were included to suggest that the study's observations may extend to different music genres.

\begin{table*}[h!]
\centering

%\tiny
%\scriptsize
%\footnotesize
\small

\begin{tabular}{|llllllll|}

 \hline
    \textbf{\#} &\textbf{Author} & \textbf{Title} & \textbf{Ref.} &  \textbf{Timing} & \textbf{Inst.} & \textbf{Struct.} &
    \textbf{Sep.} \\

\hline
   1 &  Vitalic & `No Fun' &  \citep{vitalic2005nofun} &  0'12-0'14  & Main synth. & I. & Yes \\

\hline
   2 & Vitalic & `Forgiven' &  \citep{vitalic2023forgiven} &  0'00-0'01 & Bass & QH. & Yes \\
   
\hline
    3 & Vitalic & `La Mort sur &  \citep{vitalic2012lamortsurledancefloor} & 0'03-0'04 & Bass & QH. &  No \\

    & &  le Dance Floor' & &   & & &  (solo)\\

\hline
   4 & Vitalic & `No Fun', top line,  &  \citep{vitalic2005nofun} &  0'12-0'14 & Piano & QH. & No \\

     & &  transcribed & &   & & &  (solo)\\
     & &  for piano & &   & & & \\

\hline
    5 & Vitalic & `Sweet Cigarette' &  \citep{vitalic2017sweetcigarette} &  0'29-0'31 & Main synth. & QH. & Yes \\

    & & & & & / bass & & \\

\hline
   6 & B. Tchai-& `Solo Cello Suite &  \citep{Tchaikovsky2005} &  0'15-1'38 & Cello & QH. & No  \\

    &  -kovsky &  in D Minor: III. & &   & & & (solo)\\

    &  &  Aria. Adagio.' & &   & & & \\

\hline
    7 & Vitalic & `Stamina' &  \citep{vitalic2012stamina} &  0'40-0'47 & Main synth.  & QH. & Yes \\

        & & & & & / bass & & \\

\hline
    8 & Vitalic & `Next I'm Ready' &  \citep{vitalic2012nextimready} &  0'52-0'59 & Bass & QH. & Yes \\

\hline
    9 & Vitalic & `La Mort sur &  \citep{vitalic2012lamortsurledancefloor} & 0'04-0'08 & Bass & QH. & No\\

        & &  le Dance Floor' & &   & & &  (solo)\\
   
\hline
   10 & J.S. Bach & `Partita II BWV &  \citep{Bach2010} &  1'34-1'58 & Violin & QH. & No\\

   & &   1004' in D minor: & &   & & &  (solo) \\

  & &   V. Ciaccona.' & &   & & & \\

\hline
    11 & Vitalic & `See the Sea &  \citep{vitalic2009seetheseablue} &  0'42-0'46 & Main synth. & QH. & Yes \\

        & &  (Blue)' & & & / bass & & \\

\hline
    12 &  Vitalic & `No Fun' &  \citep{vitalic2005nofun} &  0'30-0'38 & Main synth. & I. &  Yes \\

\hline

    13 &  / & / &  / &  / & Guitar & QH. & No \\

    & & & & & power chord & & \\

\hline
   
    14 &  / & / &  / &  / & 808 Woofer  & QH. & No \\

   & & & &   &  Warfare, & & \\
    & & & &   & mode 1 & & \\

\hline

\end{tabular}

\caption{List of audio samples used in the study. `I.' stands for `Inharmonic' and `QH.' for `Quasi-harmonic'.} 
    \label{table:extracts}     
    
\end{table*}

%%%%%%%%%%%%%%%%%%%%%%%%%%%%%%%%%%%%%%%%%%%%%%%%%
%% LISTENING TEST 1 - AAU
%%%%%%%%%%%%%%%%%%%%%%%%%%%%%%%%%%%%%%%%%%%%%%%%%

%\clearpage
\section{Listening Test 1: how many simultaneous pitches?}\label{sec:listening1}

\subsection{Procedure}

Listening Test 1 included samples 1 to 12. For each extract, the 59 participants reported the number of simultaneous pitches they perceived. The questionnaire collected data on age (70\% of respondents are under 35 years old), hearing condition, musical and audio expertise, as well as listening environment and monitoring system. After the test, five non-expert listeners were interviewed about their responses (see Section~\ref{sec:comments}).

Listening conditions were deliberately not controlled, in order to reflect the participants' typical listening environments. Enforcing a single set of conditions would have biased the experiment towards those specific settings. This decision is further supported \emph{a posteriori}, as Section~\ref{sec:listening2} suggests that the perceived pitches may vary depending on the listening conditions.

A preliminary experiment indicated a habituation phenomenon: the more listeners focused on the number of pitches, the more they perceived. One participant took the test twice, reporting a mean of 2.4 simultaneous pitches the first time and 3 the second time. This observation is coherent with \citepos{nunes2015power} remark, according to which lexical repetition facilitates ease of processing. To mitigate this effect, the test was offered in four versions with different extract orders. One version is accessible at: \url{https://golisten.ucd.ie/task/acr-test/6746160c75e8b306f88e34bd}.

\subsection{Results}

A preliminary Pearson correlation analysis reveals moderate associations among age, musical knowledge, and audio expertise. Age group shows a moderate positive correlation with musical knowledge ($r = 0.44$) but only a weak correlation with audio knowledge ($r = 0.19$). Meanwhile, musical knowledge and audio knowledge demonstrate a strong positive correlation ($r = 0.66$), suggesting substantial overlap between these two areas of expertise.

Table~\ref{table:extracts} shows that several samples were obtained through source separation. This is unlikely to affect the results, as the synth bass parts are highly prominent and the remaining elements of the mix consist primarily of drums, which are relatively easy to separate.

\begin{figure}[h]
\centering
\includegraphics[width=\textwidth]{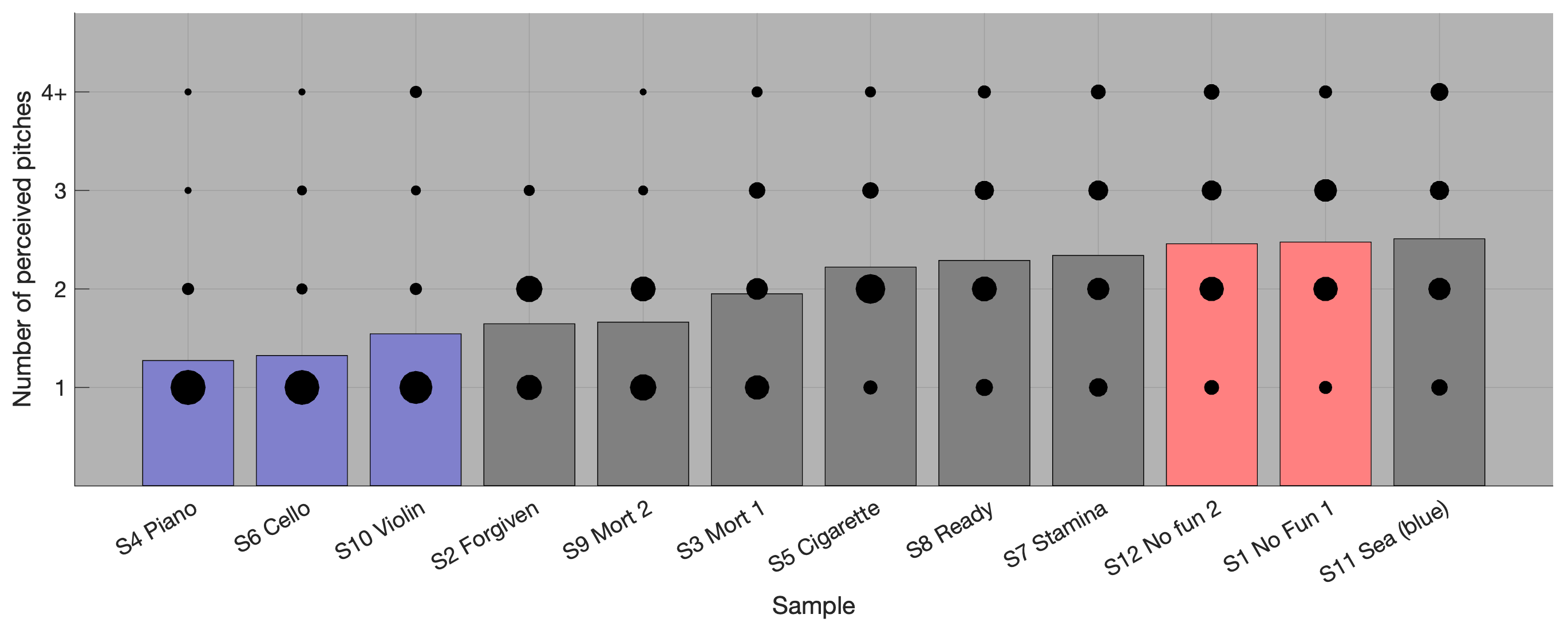}
\caption{Number of perceived pitches. See suppl. mat. Section A for the corresponding audio. Dot areas indicate the number of responses. The bar graphs represent the overall mean of the responses. The blue bars correspond to acoustic sources, and the red bars correspond to sequences of inharmonic tones.} 
\label{fig:results}
\end{figure}

Figure~\ref{fig:results} summarises the experimental results, with extracts ordered by increasing number of perceived simultaneous pitches. Not a single sample -- acoustic or electronic -- elicits a unanimous perception of a single pitch. The selected Vitalic extracts mostly lead to the perception of multiple pitches, with the mean number of perceived pitches per excerpt ranging from 1.6 to 2.5. The two samples with inharmonic tones (see Section~\ref{subsubsec:nofun}) rank higher than most, though some quasi-harmonic tones yield comparable results, and one even exceeds them.

\newpage
\textbf{Sound Categories} \quad Based on the initial expectation that acoustic instrument sounds elicit single-pitch perception, we distinguish between acoustic sounds and electronic sounds. Given that inharmonic partials may be perceived as distinct pitches \citep{bregman1994auditory,moore1985relative}, electronic sounds are further subdivided into quasi-harmonic and inharmonic categories.

\begin{figure}[h]
\centering
\includegraphics[width=1\textwidth]{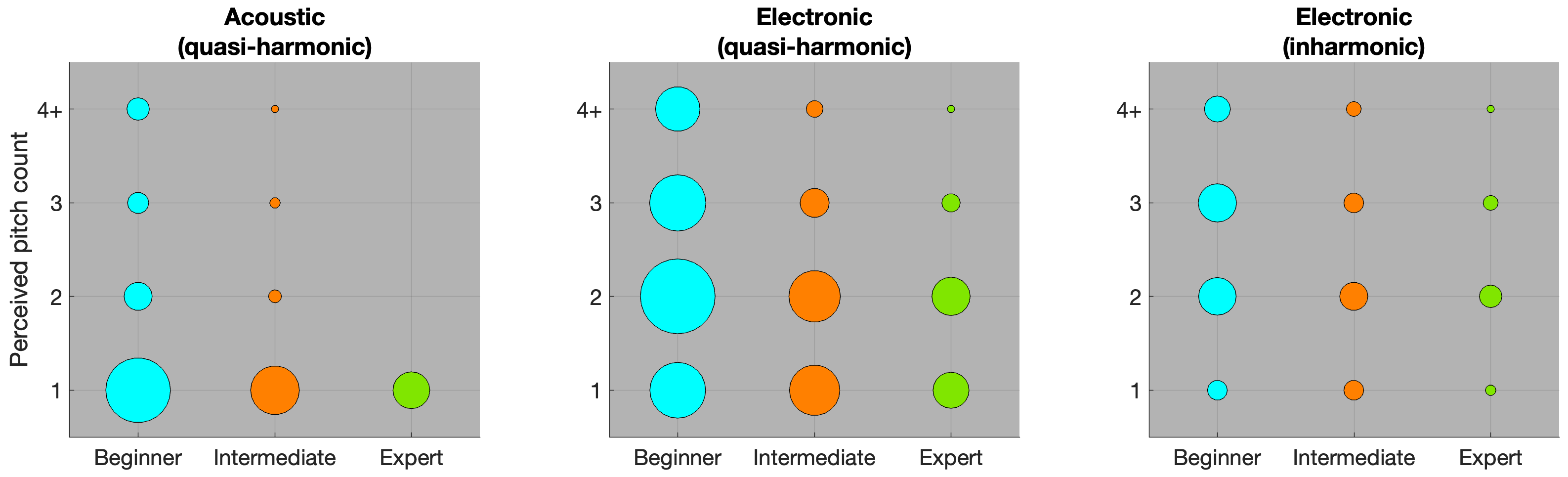}
\caption{Number of perceived pitches by musical expertise, grouped by sound category. Dot areas reflect the number of occurrences.}
\label{fig:avg-pitch-count-barplot}
\end{figure}

Figure~\ref{fig:avg-pitch-count-barplot} shows the average number of perceived pitches as a function of harmonicity and musical expertise. Both electronic categories are associated with significantly higher odds of tonal fission compared to the acoustic samples, with the strongest effect observed for the two inharmonic electronic sounds.

\vspace{.2cm}

\newpage
\textbf{Education} \quad We examine the effect of musical and audio expertise on the number of perceived pitches. As illustrated in Figure~\ref{fig:avg-pitch-count-barplot} and suppl. mat. Section~A, perceiving only a single pitch per tone occurs exclusively among audio and music experts listening to acoustic instruments.

A logistic regression analysis, treating the number of perceived pitches as an ordinal variable and controlling for age, listening device, and environment, reveals a significant negative relationship between musical expertise and pitch multiplicity ($p < 0.001$), indicating that greater musical training is associated with fewer perceived pitches. In contrast, audio expertise shows no significant effect ($p = 0.209$).

\textbf{Summary} \quad Many subjects do perceive multiple simultaneous pitches in the presented harmonic complex tones, especially for the synthetic examples. Section~\ref{sec:comments} elaborates on the subjects' perceptions of the tones.

%%%%%%%%%%%%%%%%%%%%
%% LISTENING TEST 2
%%%%%%%%%%%%%%%%%%%%

\vspace{.2cm}
\section{Listening Test 2: Transcriptions}\label{sec:listening2}
Listening Test 1 showed that quasi-harmonic complex tones may be widely perceived as transmitting multiple perceived pitches.
For Listening Test 2, we focused on tonal fission in more detail.
We selected participants with high musical expertise from the first test. These participants were asked to transcribe the pitches they heard.

\subsection{Procedure}

Seven participants transcribed the pitches they perceived in a smaller set of six samples. All participants used professional listening systems in a quiet room. The reduced number of samples compared to Listening Test 1 reflects the time-intensive nature of the task. Four samples were selected from Listening Test 1 for their diversity: one acoustic tone (sample 6), two quasi-harmonic electronic tones (samples 2 and 7), and one inharmonic tone (sample 1) for comparison. Samples 13 (power chord) and 14 (808-type bass drum) were added as representatives of specific musical genres -- rock and urban music.

In our experiment, listeners used a piano to match the pitches in the samples. They were also asked to indicate whether a tone sounded out of tune and, if so, to estimate the degree of detuning. Although the standard pitch -- matching procedure involves presenting participants with a stimulus alongside a sine wave of adjustable frequency with responses recorded as the adjusted frequency once the perceived pitches are matched \citep{shackleton1994role} -- we opted for this more familiar setup despite its lower precision, as it is more intuitive for individuals with musical training.

Listeners rated the certainty of each perceived pitch. A certainty of 1 indicates pitches that are consistently and effortlessly perceived, while low values (e.g., 0.2) reflect pitches requiring focused attention or subject to variation across listening conditions. The results were discussed both qualitatively (Section~\ref{sec:exp2-results-discussion}) and quantitatively (Section~\ref{subsec:quantitative}) and related to signal characteristics (Section~\ref{sec:signal}).

\subsection{Results}\label{sec:exp2-results-discussion}

\begin{figure}[h!]
\centering
\includegraphics[width=.95\textwidth]{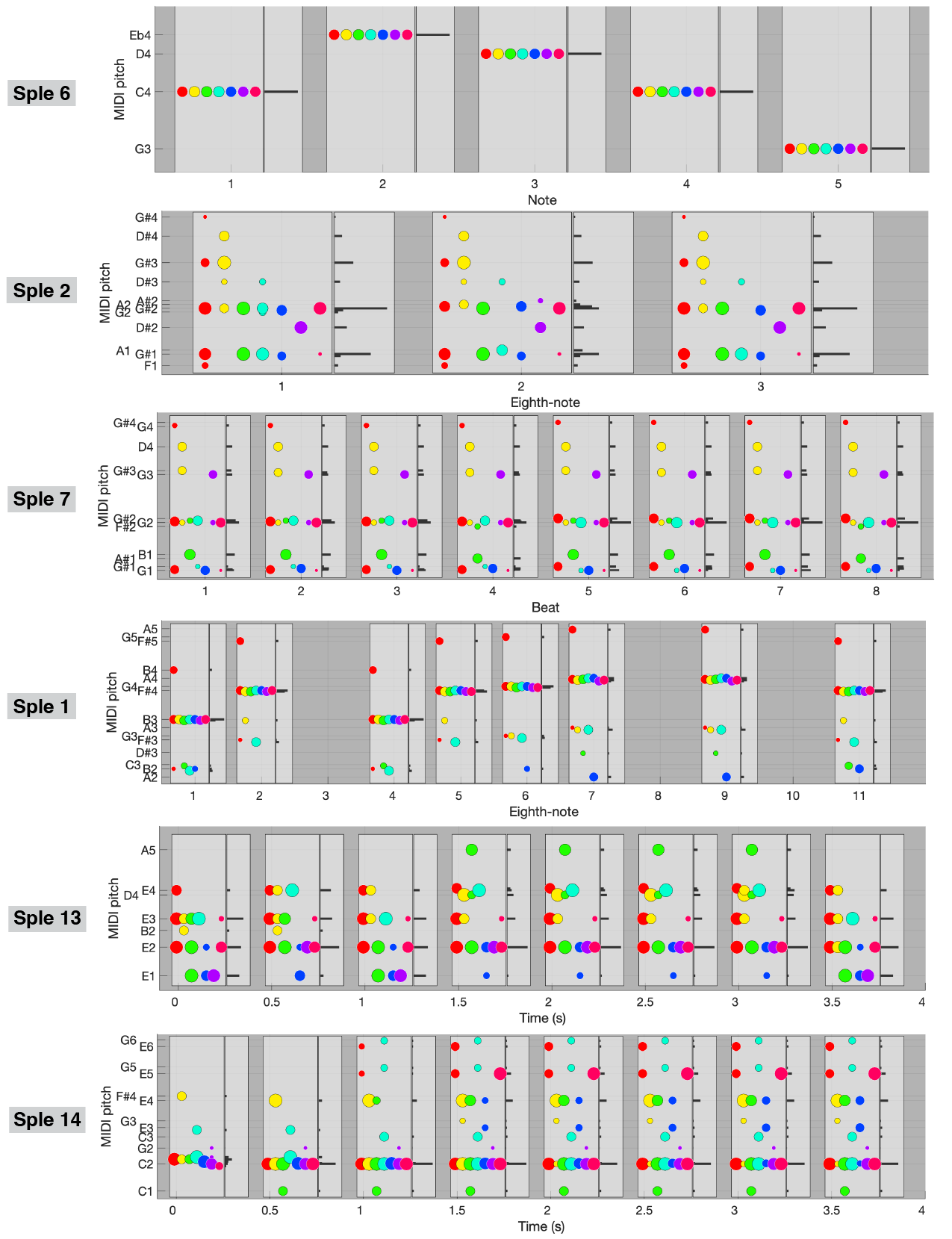}
\caption{Transcribed pitches for the six samples. Each participant is associated with one colour and one vertical position inside the clear rectangles. The area of each dot denotes certainty. The right-hand bar graphs show the distribution across subjects.} \label{fig:exp2all}
\end{figure}

See Figure~\ref{fig:exp2all} for a compilation of the transcriptions.

\vspace{.2cm}

%\label{subsec:sample6}
\textbf{Cello sequence (sample 6).} In line with the results of Listening Test~1, listeners were unanimous in their transcription of the cello example. For each note in the sequence, all listeners identified the same single pitch with maximum certainty.

\clearpage

\clearpage

\textbf{`Forgiven' (sample 2), `Stamina' (sample 7).}  Unlike the cello transcriptions, (1) listeners generally perceived multiple pitches from the tone, and (2) different listeners may perceive different pitches. These pitches are either \textit{melodically} close to the $f_0$ (around G\#1 in both cases) or \textit{harmonically} related to it in the sense of Ptolemy \citep{richter2001ptolemy}.

%\label{subsec:nofun}

\textbf{`No Fun' (sample 1).} Listener consensus on a single dominant pitch is higher than for samples 2 and 7. This contrasts with traditional views of pitch perception, where quasi-harmonic tones (samples 2 and 7) are typically associated with a single perceived pitch, while inharmonic tones (this sample) are expected to prompt multiple perceived pitches. In this sample, listeners unanimously identified a main melodic line (B3-F\#4-B3-F\#4-G4-A4-A4-F\#4). Some also perceived a parallel melodic line one octave below, while one listener reported hearing a similar line one octave above. Several listeners also perceived constant low pitches (B2).

\textbf{Power chord (sample 13).} Power chords are a defining feature of many rock styles. Although they are typically produced by simultaneously playing two or three notes on a guitar (root, fifth, and optionally the octave), the result following the distortion stage is a single harmonic complex tone -- see Section~\ref{subsec:powerchardsignal} and \citet{deruty2025vitalicnonharmonic}. For representational clarity, listeners were instructed to align their responses to a 0.5~s temporal grid during transcription. The findings exhibit several parallels with those of samples~2 and~7: multiple pitches perceived per tone, pitch perception varying between listeners, and perceived pitches that are melodically and/or harmonically related to the $f_0$ (E1).

\textbf{808 Woofer Warfare, mode 1 (sample 14).} Bass-drum tones produced by the Roland-808, a key element in many electronic and urban styles, are harmonic complex tones -- see Section~\ref{subsec:808signal} and \citet{deruty2024storch}. The results resemble those of samples 2, 7, and 14: multiple pitches per tone, listener-dependent pitch perception, perceived pitches either near or harmonically related to the expected pitch, and octave ambiguity. The range of responses is broad, spanning over five octaves. One listener perceived a C1, an octave below the $f_0$ (C2).

%\vspace{-.25cm}

\subsection{Quantitative evaluation}\label{subsec:quantitative}

This evaluation aims to determine which categories of partials the perceived pitches correspond to and to quantify the degree of listener consensus.

To assess partial categories, we measure the proportion of perceived pitches corresponding to (1) the $f_0$, (2) harmonics -- particularly octaves -- and (3) non-harmonic frequencies. In this framework, `No Fun' (sample~7) is treated as a special case due to its inharmonic tones. For this sample, the `$f_0$' is defined as the frequency spacing between consecutive partials. (see Section~\ref{subsubsec:nofun}).

To quantify consensus, we avoid using variance, as frequency differences are not inherently metric. For example, in harmonic distance (in the sense of \citet{winter2014scratch}), 200~Hz may be closer to 100~Hz than to 160~Hz. Instead, we use the Weighted Jaccard Similarity (WJS), which measures the degree to which the pitch sets -- along with their associated certainty values -- transcribed by different listeners coincide for a given tone. Let $A$ and $B$ be the sets of pitches transcribed by two listeners, and let $c_A(p)$ and $c_B(p)$ return the certainty assigned to pitch $p$ by each listener, with $c_X(p) = 0$ when $p \notin X$. The WJS is then defined as:

\vspace{.2cm}

\begin{equation}
 \label{eq:wjs}
 WJS(A, B) = \frac{\sum_{p \in A \cap B} \min ( c_A(p),c_B(p) )}{\sum_{p \in A \cup B} \max ( c_A(p),c_B(p) )}
\end{equation}

\begin{figure}[h!]
\includegraphics[width=\textwidth]{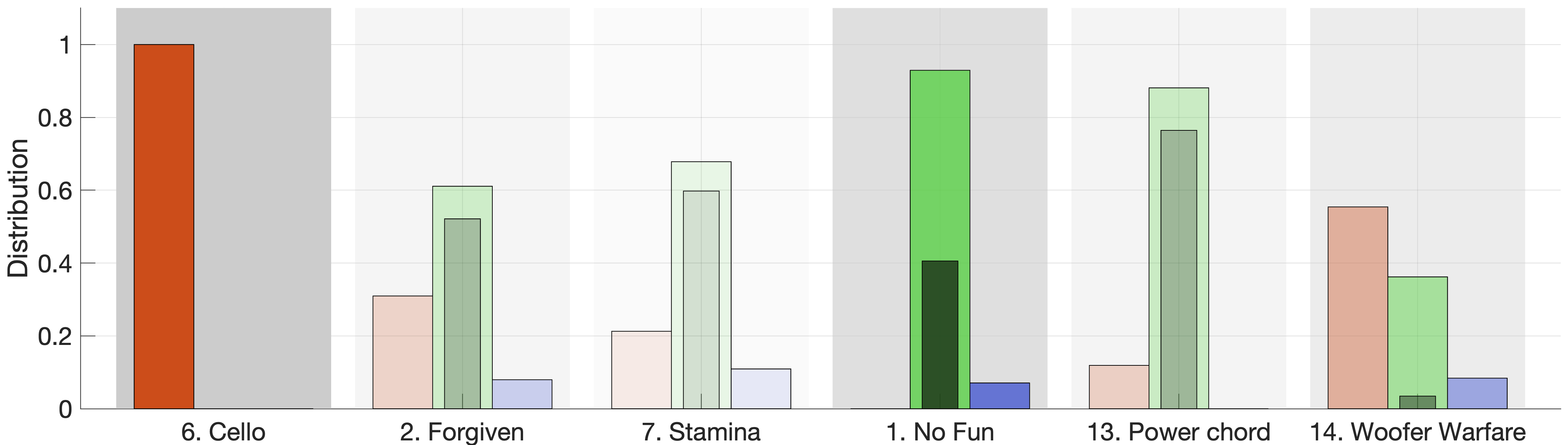}
\caption{Summary of quantitative evaluation. For each sample, the leftmost bar (red) shows the proportion of perceived pitches corresponding to the $f_0$; the middle bar (green) corresponds to harmonics, with the darker segment indicating the octaves; and the rightmost bar corresponds to other partials. The transparency of each group reflects the weighted Jaccard similarity (WJS).\\} 
\label{fig:exp2_bars}
\end{figure}

\begin{table}[h!]
\centering
\begin{tabular}{lllll}

\hline
\textbf{Sample} & \textbf{Fundamental} & \textbf{Harmonics} & \textbf{Others} & \textbf{WJS}  \\

&  & All (octaves) &  &  Mean (std. dev.) \\
\hline

Cello (sple. 6) & 1.00 & 0.00 (0.00) & 0.00 & 1.00 (0.00)  \\
Forgiven (sple. 2) & 0.31 & 0.61 (0.52) & 0.08 &  0.20 (0.08) \\
Stamina (sple. 7) & 0.21 & 0.68 (0.60) & 0.11 &  0.09 (0.01) \\
No Fun (sple. 1) & 0.00 & 0.93 (0.41) & 0.07 &  0.62 (0.10) \\
Power chord (sple. 13) & 0.12 & 0.88 (0.76) & 0.00 &  0.22 (0.04) \\
808 Woofer Warfare (sple. 14) & 0.55 & 0.36 (0.03) & 0.08 & 0.38 (0.16)  \\

\hline

\end{tabular}

\caption{Value outputs for the evaluation criteria.  Proportion of perceived pitches corresponding to the fundamental, to harmonics (of which octaves), and to other partials. Weighted Jaccard similarity.}
\label{table:values}  

\end{table}

\newpage

Table~\ref{table:values} presents the evaluation results, while Figure~\ref{fig:exp2_bars} provides a visual representation. Several observations can be made:

\begin{itemize}
    \item For sample~6 (cello), all values indicate unanimous pitch perception across listeners, suggesting a clear and stable pitch is conveyed.
    
    \item Samples~2 and~7 (synthesizer, quasi-harmonic) produce similar outcomes: a lack of consensus among listeners, with most perceived pitches corresponding to octaves of the $f_0$. This suggests that pitch ambiguity is concentrated around octave-related partials.
    
    \item Sample~7, which involves inharmonic tones, surprisingly yields higher consensus than the non-acoustic harmonic tones. The pitch corresponding to the difference between partials -- potentially perceivable through temporal modeling -- is never reported. However, upper octaves account for a substantial proportion of the perceived pitches. The prevalence of harmonics in listener responses may be partly explained by the high density of partials in the register where these pitches occur.

    \newpage
    
    \item In sample~13 (power chord), the majority of perceived pitches derive from upper octaves.
    
    \item Sample~14 stands out among non-acoustic quasi-harmonic tones. It exhibits strong perception of the fundamental frequency and very low octave ambiguity. Rather than the octave, other harmonics appear to be more salient in listeners' perceptions.
    
\end{itemize}

%\newpage

\section{Listening Test 2: results and sample signals}\label{sec:signal}

The contrast between the extent of tonal fission observed in Listening Test 2 and the common association of harmonic complex tones with a single pitch calls for a closer look at the samples to explore potential links between this phenomenon and signal features.

\begin{figure}[h!]
\includegraphics[width=\textwidth]{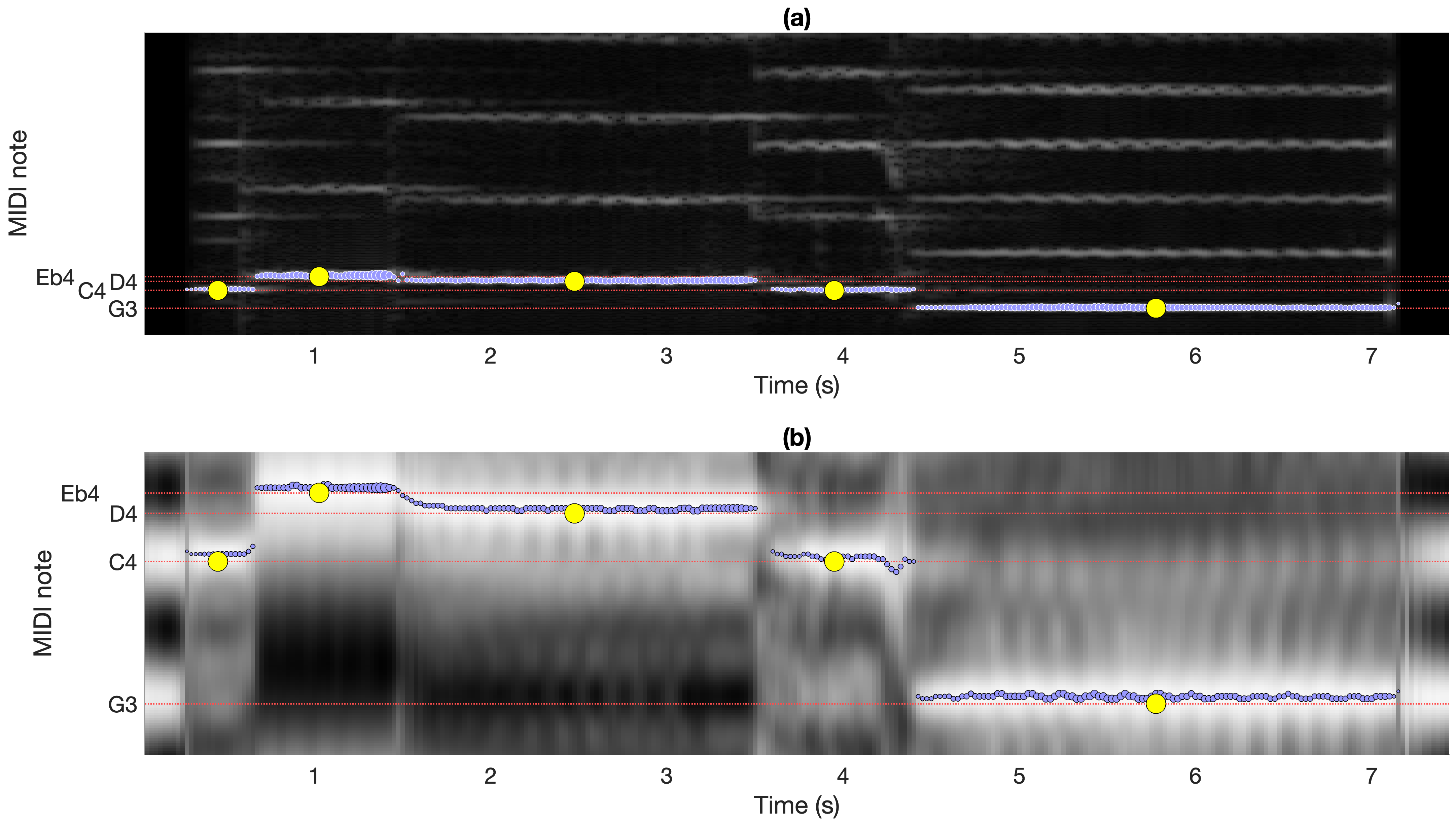}
\caption{Cello (sample~6). (a) STFT, (b) STAC. See Suppl. mat., Section~B, for the same figure with synchronized audio. The yellow scatter plots indicate the listener transcriptions shown in Figure~\ref{fig:exp2all}; the blue scatter plots indicate the maxima for each frame.}
\label{fig:CelloCorr}
\end{figure}

\subsection{Cello sequence (sample 6)}

Figure~\ref{fig:CelloCorr} shows the STFT (Short-Term Fourier Transform) and STAC (Short-Term Autocorrelation) of sample~6, illustrating a case in which tonal fission is not observed. The $f_0$ and the first peak of the autocorrelation function after the zero crossing align and correspond to the perceived pitches. Both spectral and temporal modeling lead to the same pitch. These observations are consistent with established findings on pitch perception in harmonic complex tones (e.g., \citet{goldstein1973optimum, licklider1951duplex, plomp1967pitch, de2002yin}).

\subsection{`Forgiven' (sample 2)}

\begin{figure}[h!]
\centering
\includegraphics[width=\textwidth]{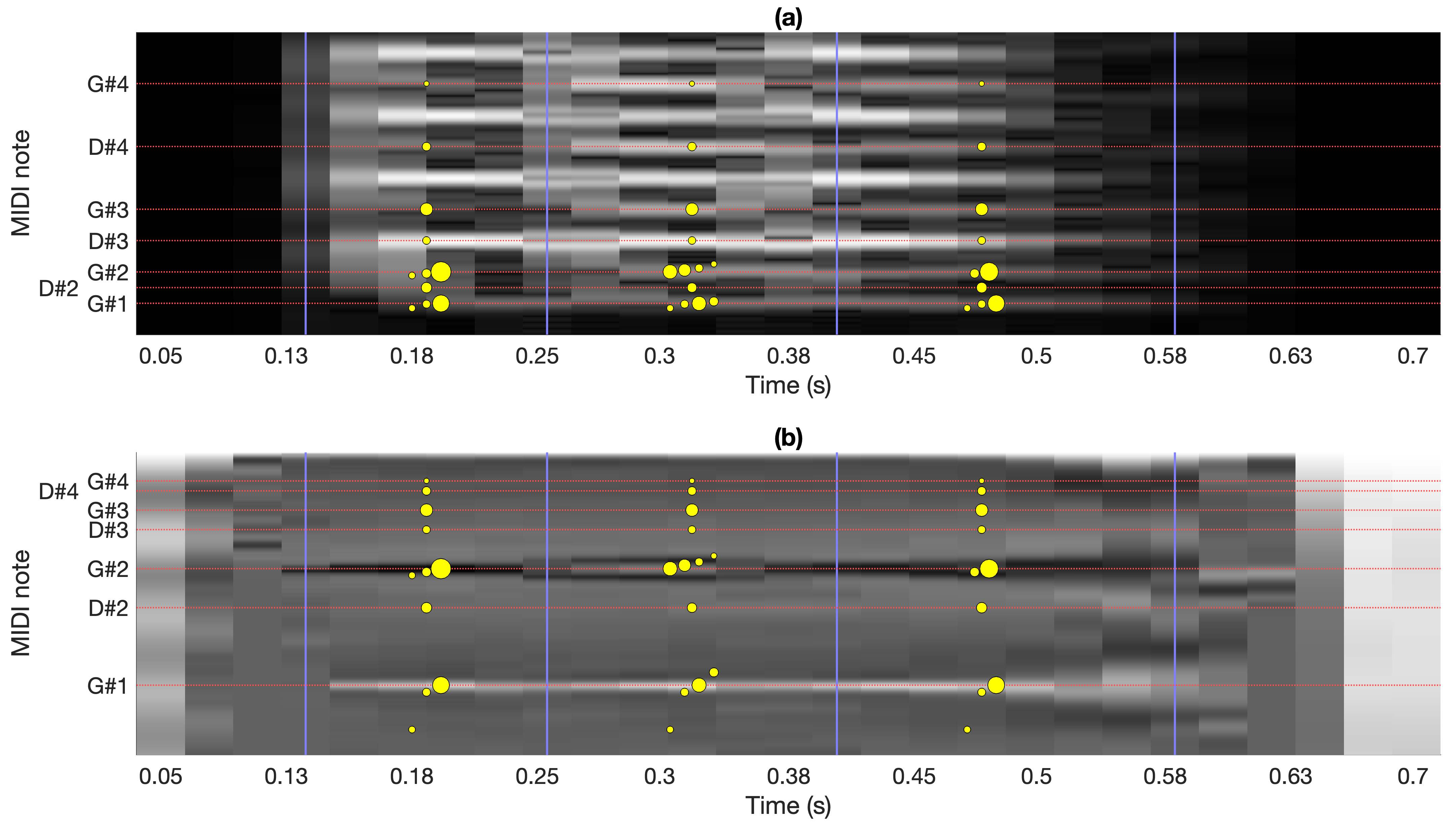}
\caption{`Forgiven', bass, 0'00--0'01 (sample 2), weighted audio. See suppl. mat. Section C for the same figure with synchronized audio. (a) Short-Time Fourier Transform (STFT), (b) Short-time Auto Correlation (STAC). The vertical blue lines indicate note boundaries.} \label{fig:ForgivenCorr}
\end{figure}

\textbf{Sample details.} Sample 2 was produced using Xferrecords' Serum, a software wavetable synthesiser. As shown in Figure~\ref{fig:ForgivenCorr}(a), the sample consists of three consecutive quasi-harmonic tones with an $f_0$ near G$\sharp$1. All partial frequencies decrease slightly and gradually throughout each tone.

\textbf{Perceived pitches and signal.} Figure~\ref{fig:ForgivenCorr}(a) shows that most perceived pitches align with partials, suggesting spectral modeling. G$\sharp$1 ($f_0$) and G$\sharp$2 (harmonic~1) are also consistent with temporal modeling. Indeed, the first and third tones primarily contain odd harmonics and exhibit a relatively weak fundamental, resembling harmonic complexes with missing lower partials, as described by \citet{yost2009pitch}. In such cases, pitch perception becomes ambiguous -- corresponding either to the $f_0$ (see Figure~\ref{fig:ForgivenCorr}(b), autocorrelation peak at G$\sharp$1) or to the spacing between partials. This spacing is reflected in the first minimum of the autocorrelation function following the initial peak \citep{deruty2025vitalictemperament}, which in Figure~\ref{fig:ForgivenCorr}(b) corresponds to G$\sharp$2. One perceived pitch remains unexplained: D$\sharp$2, approximately one octave below the fifth harmonic.

\vspace{.5cm}

\newpage

\subsection{Stamina (sample 7)}

\begin{figure}[h!]
\centering
\includegraphics[width=\textwidth]{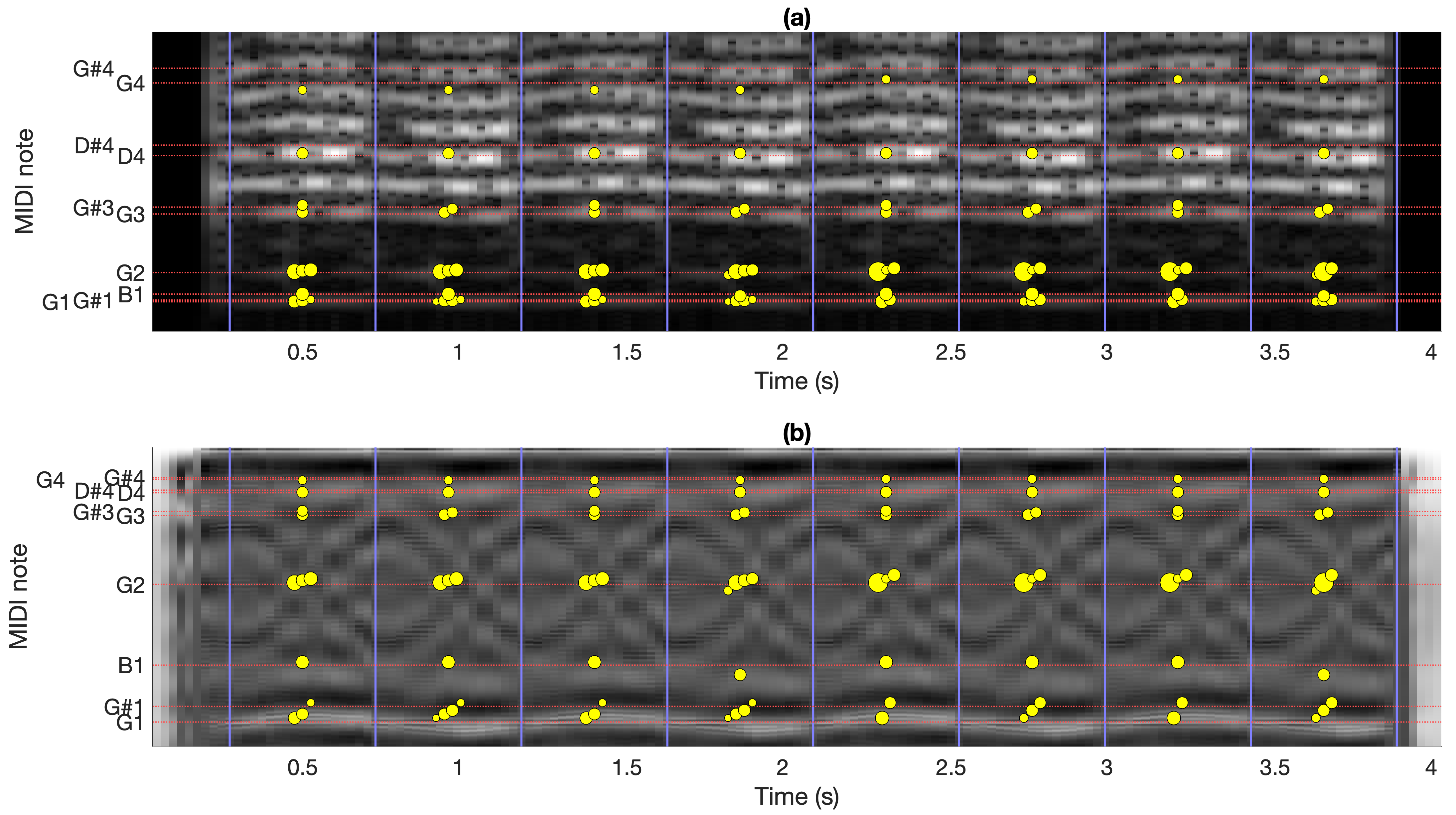}
\caption{`Stamina', main synthesiser/bass, 0'40-0'47 (sample 7), weighted audio. See suppl. mat. Section D for the same figure with synchronized audio. (a) STFT, (b) STAC. The vertical blue lines show the `note' start.} \label{fig:StaminaCorr}
\end{figure}

\textbf{Sample details.} Sample 7 was created using a Novation Bass Station, a 1993 analogue synthesizer. It features a quasi-harmonic tone with an $f_0$ near G1 and attenuated partials above the fifth harmonic. Partial frequencies are modulated conjointly, while amplitudes are modulated individually. Above the fourth partial, odd and even harmonics alternate. As shown in Figure~\ref{fig:StaminaCorr}(b), modulation of partial frequencies produces several closely spaced autocorrelation peaks that evolve periodically.

\textbf{Perceived pitches and signal.} Figure~\ref{fig:StaminaCorr}(a) illustrates the relatively low loudness of the lower harmonics, suggesting that the perception of low pitches (around G1 and G2) relies on temporal modeling. In contrast, the alignment between higher perceived pitches and upper partials indicates a reliance on spectral modeling. The continuous modulation of partial frequencies may account for the variety of perceived tunings.

\vspace{.2cm}

\subsection{`No Fun' (sample 1)}\label{subsubsec:nofun}

\textbf{Sample details.} The main synthesiser part in `No Fun' was built from tones sampled from the Roland VariOS unit. It consists of a sequence of inharmonic tones exhibiting what \citet{yost2009pitch} describes as `pitch-shift of the residue'. \citet{deruty2025vitalicnonharmonic} provide a detailed analysis of the sample. The frequency differences between adjacent partials remain approximately constant throughout the extract (around B1), even though the partials themselves are shifted.

\begin{figure}[h!]
\centering
\includegraphics[width=\textwidth]{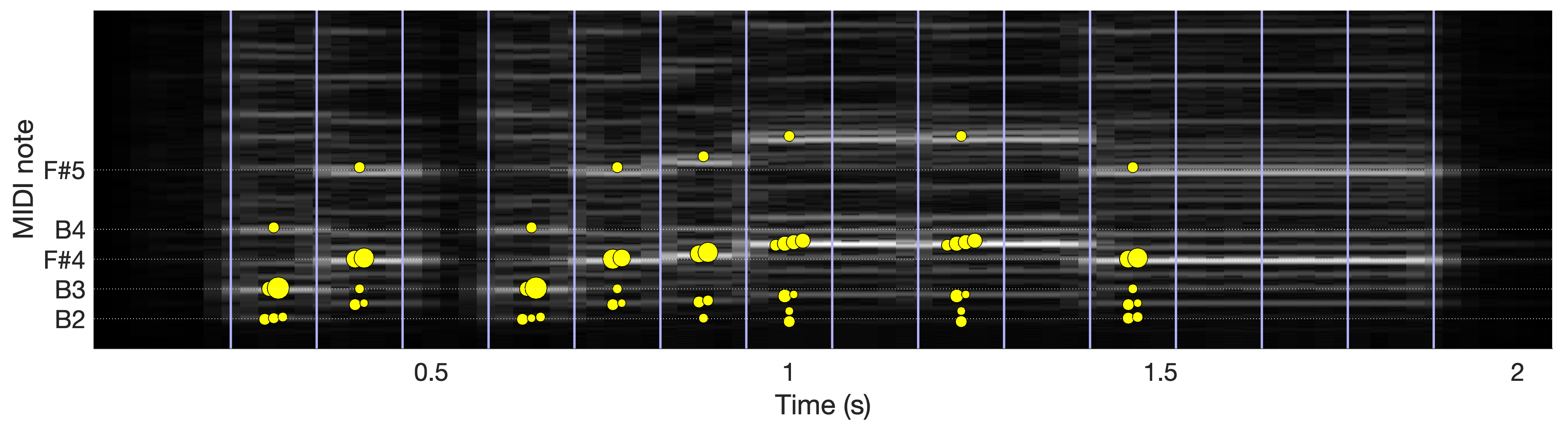}
\caption{`No Fun', 0'12 to 0'14, main synthesiser part (sample 1). See suppl. mat. Section E for the same figure with synchronized audio. STFT, weighted audio. The vertical blue lines indicate the positions of the eighth-notes.} \label{fig:NoFunSTFT}
\end{figure}

\newpage
\textbf{Perceived pitches and signal.} Figure~\ref{fig:NoFunSTFT} overlays the annotation results with the sample's STFT. STAC is not included, as it is not suited for pitch detection in inharmonic tones \citep{deruty2025vitalictemperament}. Pitch perception appears to be primarily driven by spectral modeling, with most perceived pitches corresponding to prominent partials. Although low sensitivity to bass frequencies may explain the absence of B1, it does not account for the perception of B2 in its place. Temporal modeling may explain why a B is perceived, yet it remains unclear why the upper octave is heard rather than the frequency difference itself.

\vspace{.3cm}

\begin{figure}[h!]
\centering
\includegraphics[width=\textwidth]{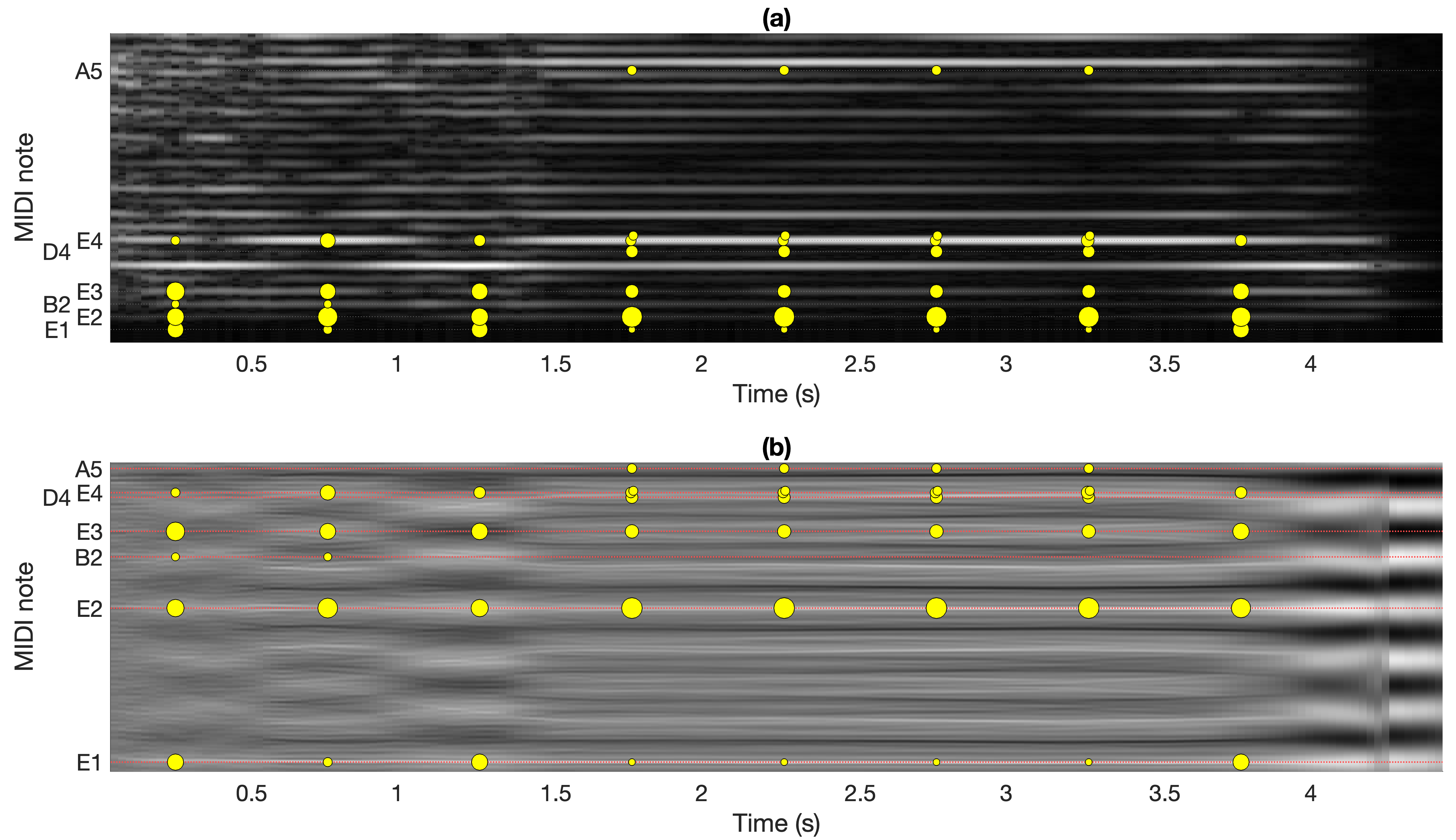}
\caption{Power chord (sample 13), weighted audio. See suppl. mat. Section F for the same figure with synchronized audio. (a) STFT, (b) STAC.} \label{fig:PowerChordsSignal}
\end{figure}

\subsection{Power chord (sample 13)}\label{subsec:powerchardsignal}

\textbf{Sample details.} The played notes are E2, B2, and E3. The guitar is processed with Native Instrument's Guitar Rig `Rammfire' amp emulation. Distortion applies a non-linear transformation to the signal, inducing intermodulation distortion.  Intermodulation distortion generates partials from each input partial and their combinations \citep{newell2017recording}, resulting in a quasi-harmonic complex tone with E1 as $f_0$. The partial corresponding to the $f_0$ is weak.

\textbf{Perceived pitches and signal.} Figure~\ref{fig:PowerChordsSignal}(a) suggests that, as in sample~7, the relatively low loudness of the lower harmonics leads to the perception of low pitches (E1, E2) primarily through temporal modeling. Figure~\ref{fig:PowerChordsSignal}(b) further indicates that the preference for E2 may be related to a more clearly defined autocorrelation peak. In contrast, the alignment of higher perceived pitches (D4, E4, A5) with upper partials supports spectral modeling. The variation in perceived pitch resulting from changes in partial energy distribution underscores the role of upper partial loudness in facilitating tonal fission.

\newpage
\subsection{808 Woofer Warfare (sample 14)}\label{subsec:808signal}

\begin{figure}[h!]
\centering
\includegraphics[width=\textwidth]{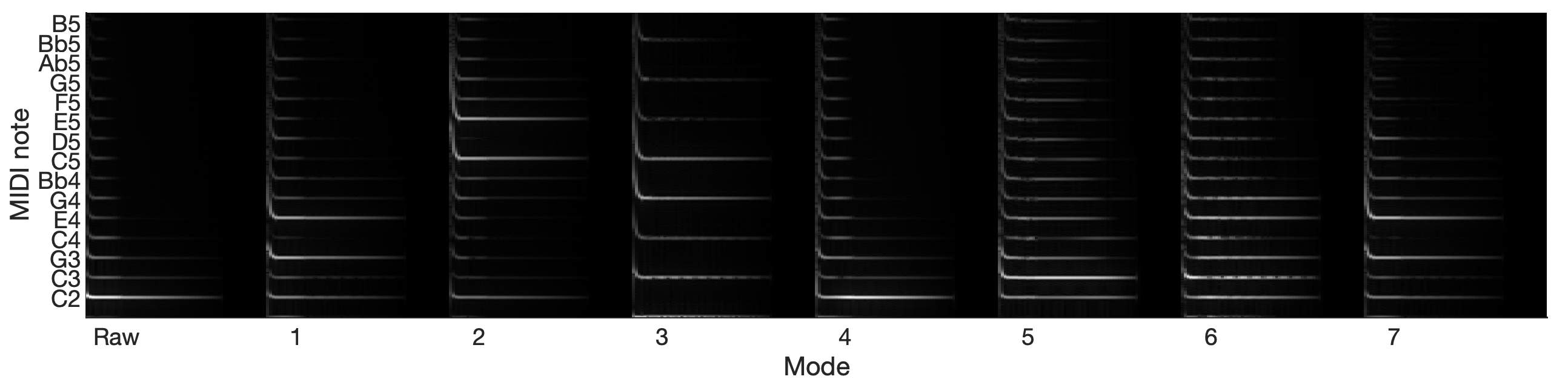}
\caption{Omnisphere, Seismic Shock library, `808 Woofer Warfare' patch, STFT for the seven modes. Unweighted audio. See suppl. mat. Section G for the same figure with synchronized audio.} \label{fig:808presets}
\end{figure}

\textbf{Sample details.} The sample was produced using the Seismic Shock library in Omnisphere. The `808 Woofer Warfare' patch, designed by Sonic Extensions, features `tone modes' -- presets that shape the spectrum of an 808 kick drum sample. Figure~\ref{fig:808presets} shows the STFT of the original sample alongside its seven tone modes. Like the original, each mode begins with a continuous frequency drop, with the $f_0$ settling at C2. Sample~14 uses mode~1, which emphasizes harmonic~5 (E4), a mode favored by the Hyper Music company in production \citep{deruty2025primaal}.

\begin{figure}[h!]
\centering
\includegraphics[width=\textwidth]{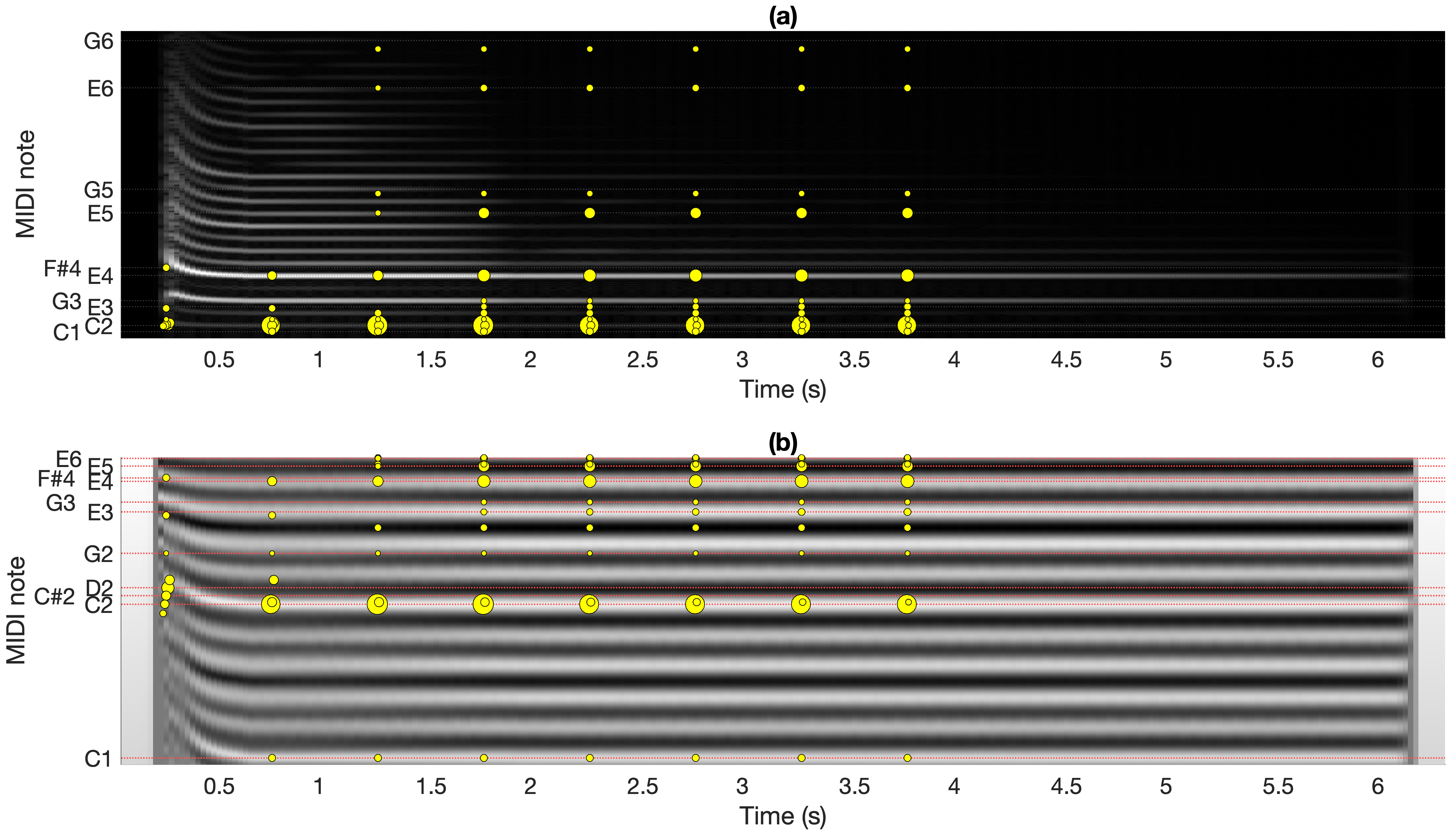}
\caption{808 Woofer Warfare, mode 1 (sample 14), weighted audio. See suppl. mat. Section G for the same figure with synchronized audio. (a) STFT, (b) STAC.} \label{fig:WooferWarfareSignal}
\end{figure}

\textbf{Perceived pitches and signal.} As in samples 7 and 13, the STFT, here in Figure~\ref{fig:WooferWarfareSignal}(a), suggests an association between higher perceived pitches and spectral modeling, while the low energy of the lower partials supports the perception of low pitches (C1, C2) via temporal modeling. The distinct autocorrelation peak in Figure~\ref{fig:WooferWarfareSignal}(b) further reinforces the link between C2 and temporal modeling.

\clearpage

\vspace{-.4cm}

\vspace{-.4cm}
\subsection{Signal analysis: conclusion}

Signal analysis of the Listening Test 2 samples suggests that tonal fission in non-acoustic quasi-harmonic tones (samples~3, 7, 13, and~14) may arise from (1) spectral modeling of loud upper partials and (2) temporal modeling that leads to different pitch outcomes than spectral modeling. Typically, higher-frequency partials are associated with spectral modeling, while lower-frequency partials are linked to temporal modeling.

Similar observations apply to the electronic inharmonic tone (sample~1), but not to the instrumental quasi-harmonic tone (sample~6). The next section proposes that the exceptional status of sample~6 may be due not only to its intrinsic acoustic properties but also to cultural conditioning.

\clearpage

%%%%%%%%%%%%%%%%%%%%%%%%%%%%%%%%%%%%%%%%%%%%%%%%%
%% COMMENTS
%%%%%%%%%%%%%%%%%%%%%%%%%%%%%%%%%%%%%%%%%%%%%%%%%

\section{Interpretation of results through listener comments}\label{sec:comments}

This section synthesizes feedback from participants in both listening tests, highlighting inter-listener variability, the influence of listening conditions, and the ambiguity between pitch and timbre. These findings challenge the reliability of absolute pitch descriptors and support the hypothesis that pitch perception may, in some cases, be multistable.

\subsection{Comments from expert listeners (Listening Test 2)}\label{subsec:expertcomments}

Expert listeners (EL) had the following backgrounds: EL~1, 2, 4, and~6 were familiar with both Western classical music (WCM) and contemporary popular music (CPM); EL~3 had a background in CPM; EL~5 in WCM and psychoacoustics; and EL~7 in WCM.

\vspace{.2cm}

\textbf{Inter-listener variability.} Expert listeners diverged in how they interpreted the tones. EL~6 and EL~7, both with classical backgrounds, tended to associate each tone with a single pitch, sometimes relegating additional pitches to the realm of timbral texture. In contrast, EL~3, experienced in rap production, regarded multi-pitch perception as a normative aspect of tone design. In the case of acoustic tones, EL~4 noted the distinction between reporting what is actually heard (multiple pitches) and what is inferred to have been played (a single note).

\textbf{Effect of listening conditions.} Perceived pitch varied notably with playback conditions. EL~1--4 and EL~6 reported changes in pitch perception depending on the monitoring system, listening position, and playback level. These effects align with known variations in frequency response across speaker types \citep{newell2001yamaha}, listener angles \citep{evans2009effects}, and loudness curves \citep{iso2262023}. While pitch perception in acoustic samples (cello) remained stable, synthetic tones were sensitive to these variables.

\textbf{Perceptual flexibility.} Three listeners noted that, even under fixed listening conditions, repeated exposure allowed them to shift attention between different pitches -- effectively choosing which ones to perceive within a limited set. This perceptual flexibility is further discussed in Section~\ref{subsec:Necker}.

\textbf{Pitch--timbre ambiguity.} All expert listeners found it difficult to disentangle pitch from timbre in non-acoustic tones. In some cases, even though pitches were perceived, they were interpreted as timbral features instead of distinct melodic elements. EL~7, for example, reported hearing multiple pitches but \textit{chose} to interpret them as part of the tone's character rather than as discrete entities.

\textbf{Pitch identification in the lower register.} Several listeners, notably EL~4 and EL~7, highlighted challenges in identifying precise pitch values in the lower frequency range (MIDI octaves~1 and~2). They attributed this difficulty both to the perceptual characteristics of low tones -- specifically, their reduced pitch strength, as noted by \citet{huron2001tone} -- and to limited exposure to such registers in classical and earlier popular music \citep{deruty2024storch, hove2019increased}.

\textbf{Octave ambiguity in the lower register.} EL~3 remarked that in rap production, `bass' sounds are not expected to convey a specific octave; a single keypress may trigger energy across two octaves, consistent with prior observations \citep{deruty2025primaal}. EL~5 and EL~7 reported that octaves in the lower register often merged perceptually into a single stream in the sense of \citet{bregman1994auditory}. EL~5 compared this to the effect of parallel octaves in Romantic piano music.

\textbf{Effects of non-12-TET tuning.} EL~1 and EL~2 observed that deviations from 12-tone equal temperament, as documented in Vitalic's music \citep{deruty2025vitalictemperament}, complicated the task of assigning pitch names. They remarked that such deviations made it more difficult to identify independent lines.

\subsection{Comments from non-expert listeners (Listening Test 1)}\label{subsec:nonexpertcomments}

\textbf{Inter-listener variability.} Unlike experts, non-expert listeners described most Vitalic samples in consistent terms. They typically identified distinct `low' and `high' layers and considered high-frequency buzzing components (e.g., in samples~1, 7, 9, and 12) to be melodic. They found it difficult to count the number of simultaneous pitches during rapid changes and noted that masking between layers made this task harder.

\textbf{Perception of low-frequency melodies.} Unlike experts, who often struggled to assign clear pitches in the low register, two non-expert listeners reported that low-frequency melodic lines were easier to follow than other lines, as they were well separated from other elements.

\textbf{Perception of upper harmonics in acoustic sources.} Two non-expert participants perceived a second melodic layer in the cello sample, attributing it to overtone richness. That only non-experts mentioned this perception runs counter to Mersenne's historical claim that only the `best ears' can detect upper harmonics \citep{mersenne1636harmonie}.

\textbf{Violin reverberation and perceived simultaneity.} In the violin sample (sample~10), reverberation led one non-expert to report two simultaneous pitches, while others hesitated. This reflects a gap between the number of acoustically present tones and what is perceptually interpreted as distinct pitches. Indeed, expert listener 4 noted that while multiple pitches were always present, they were perceived as simultaneous only in short (sub-second) isolated extracts. Refer to suppl. mat., Section H, for a video showing that multiple tones are indeed physically present due to reverberation, even if not always perceived as simultaneous.

\subsection{Interpretation of listener comments}

The listener comments suggest a fragmented and context-dependent landscape of pitch perception.

\begin{itemize}
    \item Experts varied in their pitch interpretation strategies, often influenced by musical background. Those trained in classical music tended toward single-pitch reporting, while others embraced tonal fission and ambiguity as stylistically appropriate.
    
    \item Pitch perception was especially unstable in the lower register, where octave ambiguity and tonal fusion frequently obscured register identification.
    
    \item Listening conditions -- system, position, and level -- significantly impacted pitch perception in synthetic tones.
    
    \item Several expert listeners demonstrated perceptual flexibility, suggesting that pitch attention can be selectively directed, not just passively received.
    
    \item The boundary between pitch and timbre was blurred in complex tones, often preventing clear categorization.
    
    \item Non-expert listeners tended to report more consistent perceptions across samples. They identified clear high and low layers, on some occasions showing sensitivity to melodic content that experts overlooked.
    
    \item The perception of upper harmonics in monophonic instruments by non-experts challenges assumptions about auditory expertise.
    
\end{itemize}

Taken together, these findings confirm that pitch should \textit{not} be treated as an objective, intrinsic property of sound, but as a perceptual construct -- shaped by listening context, cultural familiarity, and individual focus. Section~\ref{subsec:Necker} elaborates on the multistable aspect of pitch, while Section~\ref{subsec:rethink} discusses implications for theoretical models.

%%%%%%%%%%%%%%%%%%%%%%%%%%%%%%%%%%%%%%%%%%%%%%%%%%%%%%%%%%%%%%%%%%%%%%%%%%%%%%%%%%%%%%%%%%%%%%%%%%%%%%%%%%%%%%%%%%%%%%%%%%%%

\section{Discussion}\label{sec:discussion}

\subsection{Rethinking pitch in perception and music analysis}\label{subsec:rethink}

Western classical music is grounded in notated scores that encode properties such as pitch. Instruments in this tradition typically produce harmonic or quasi-harmonic tones via acoustic resonance, with the perceived pitch usually aligning with the fundamental frequency ($f_0$). This framework has shaped key tasks in Music Information Retrieval (MIR), such as pitch tracking \citep{drugman2018traditional,kim2018crepe} and audio-to-MIDI alignment, where MIDI files serve as symbolic representations \citep{ewert2012towards,benetos2018automatic}. The reliance on notation in analysis has led to what \citet{middleton1990studying} calls the `reification of the score', treating it as an idealized version of the music.

This model assumes that pitch is an ontological property of a tone -- that each tone corresponds to a single, context- and listener-independent pitch. However, such assumptions may be challenged. Prior research \citep{deruty2025vitalictemperament,deruty2025primaal}, this study, and the producer's \emph{deliberate} use of pitch uncertainty suggest that these views may be inadequate for musical contexts beyond Western classical music.

Artists like Vitalic and the Hyper Music producers \citep{deruty2025primaal} deliberately manipulate pitch perception by varying the number of pitches per tone and introducing pitch ambiguity as a compositional parameter. The commercial success of their work points to an alternative paradigm, in which tones may be designed to transmit not fixed pitches but sets of possible ones. The use of power chords in rock and metal further indicates that this approach extends beyond electronic music into other popular genres.

This shift prompts a redefinition of what constitutes a musical tone. Rather than equating a tone with a single perceived pitch \citep{moore1986thresholds}, an alternative view defines it as a perceptual stream of partials that share a \textit{common fate} \citep{rasch1982perception,Wertheimer1938}, with some pitches being more likely to be perceived than others. This definition is adopted in Section~\ref{subsec:terminology} and in \citet{deruty2025vitalictemperament}. From this perspective, pitch may be described not as a fixed value but as a probability distribution. A related interpretation treats the probability of perceiving a given pitch as a measure of pitch strength \citep{deruty2024pitchstrength}.

\subsection{The musical role of multi-stable perception}\label{subsec:Necker}

Multistability arises when a single physical stimulus produces alternating subjective percepts \citep{schwartz2012multistability}. A classic example is the `Necker cube' \citep{necker1832lxi}, a two-dimensional drawing that can be seen with either the lower-left or upper-right square as the front face. In audio, multistability (in this case, bistability) includes: (1) the tritone paradox, where a pair of Shepard tones \citep{shepard1964circularity} separated by a tritone is heard as ascending or descending \citep{deutsch1991tritone}, and (2) an A-B tone sequence heard as one stream (ABA-ABA) or two: A-A-A-A and -B-B-B- \citep{pressnitzer2006temporal}.

Section~\ref{subsec:expertcomments} notes that listeners in test 2 could deliberately \textit{choose} one pitch set over another, indicating multistable perception -- suggesting quasi-harmonic complex tones can yield multistable percepts. Constructing tones as multistable may help sustain listener interest. \citet{madison2017repeated} found that musical appreciation declines with repeated exposure in a single session. A similar pattern appears in Vitalic's tracks, which often feature heavy repetition (see suppl. mat., Section C).  \citet{orbach1963reversibility} explain Necker cube reversals via `satiation of orientation': once a perceptual process is saturated, a switch occurs. This suggests that producers may use multistable tones to enable perceptual switching, delaying satiation and maintaining appeal across repeated hearings. These hypotheses warrant further study.

%% %%%%%%%%%
%% Conclusion
%% %%%%%%%%%%%

\section{Conclusion}\label{sec:conclusion}

The electronic music producer Vitalic deliberately designs quasi-harmonic tones that convey multiple simultaneous pitches -- a phenomenon we refer to as \textit{tonal fission}. Building on prior studies that examined the signal properties of such tones in the music of Vitalic and others \citep{deruty2025vitalictemperament,deruty2025primaal}, the present study employed listening tests to further explore this perceptual effect.

Listening Test~1 investigated the number of simultaneous pitches perceived in electronic and acoustic tones within a musical context. Results indicate that quasi-harmonic tones -- particularly electronic ones -- can convey multiple pitches simultaneously, with perceptual outcomes varying across listeners.

In Listening Test~2 (Section~\ref{sec:listening2}), musically trained participants transcribed the pitches they perceived in tone sequences. These transcriptions confirmed that, apart from the acoustic tone sequence, multiple pitches were often perceived simultaneously, with the specific pitch identifications varying across participants. Signal analysis suggests that this multiplicity of perceived pitches (tonal fission) may result from spectral modeling of prominent partials, potentially conflicting with the outcomes of temporal modeling.

%both temporal and spectral modeling of the signal. %, including upper harmonics.

Section~\ref{sec:comments} presented listener comments that emphasized inter-listener variability, the influence of listening conditions, and the ambiguity between pitch and timbre perception. These findings underscore the contextual and subjective nature of pitch perception in electronically produced music and suggest that tones may serve different functions in popular music -- transmitting a set of pitches that may or may not be perceived -- than in Western classical traditions, where (typically) a single intended pitch is expected to be perceived corresponding to each note.

Section~\ref{sec:discussion} explored broader implications, including challenges to conventional definitions of musical tones and to pitch-tracking methods in music analysis. It also addressed the relevance of multistable perception in quasi-harmonic tones.

Pitch may not be an intrinsic property of a sound object, but rather a listener-constructed, context-dependent percept. A single harmonic tone in a musical context can be deliberately designed to project multiple, ambiguous pitches to listeners through tonal fission. This challenges the definition of a `note` in contemporary popular music.

These conclusions open several directions for future research: for instance, better examining the cognitive mechanisms underlying tonal fission; exploring its pervasiveness in a broader range of musical contexts; further investigating the potential transition from active to resultant pitch in contemporary popular music; and confirming or refuting the existence of multi-stable pitch perception in this context.

\vspace{.2cm}
\section{Acknowledgments}

Many thanks to: Cyran Aouameur, Amaury Delort, Matthias Demoucron, Stefan Lattner, Sony Computer Science Laboratories, Paris, France. Michael Turbot, Sony A.I., Z{\"u}rich, Switzerland. Yann Mac\'e and Luc Leroy from the music production company Hyper Music.

\clearpage

\bibliographystyle{apalike}
\bibliography{Vitalic2arxiv.bib}
\end{document}